\documentclass[12pt]{iopart}
\usepackage{iopams}  
\usepackage{bm}
\usepackage{xcolor}
\usepackage{mathrsfs}
\usepackage[utf8]{inputenc}
\usepackage{enumerate}
\usepackage{appendix}
\usepackage{dcolumn}
\usepackage{multirow}
\usepackage{float}

\begin{document}

\title{Essential core of the Hawking--Ellis types}

\author{Prado Mart\'{\i}n--Moruno$^1$ and Matt Visser$^2$}

\address{$^1$Departamento de F\'isica Te\'orica and UPARCOS, \\Universidad Complutense de Madrid,
E-28040 Madrid, Spain}
\address{$^2$School of Mathematics and Statistics, Victoria University of Wellington, \\
PO Box 600, Wellington 6140, New Zealand}
\ead{pradomm@ucm.es, matt.visser@sms.vuw.ac.nz}
\vspace{10pt}

\begin{abstract}
The Hawking--Ellis (Segre--Pleba\'nski) classification of possible stress-energy tensors is an essential tool in analyzing the implications of the Einstein field equations in a more-or-less model-independent manner. 
In the current article the basic idea is to simplify the Hawking--Ellis type I, II, III, and IV classification by isolating the ``essential core'' of the type II, type III, and type IV stress-energy tensors; this being done by subtracting (special cases of) type I to simplify the (Lorentz invariant) eigenvalue structure as much as possible without disturbing the eigenvector structure.
We will denote these ``simplified cores'' type II$_0$, type III$_0$, and type IV$_0$.
These ``simplified cores'' have very nice and simple algebraic properties.
Furthermore, types I and II$_0$ have very simple classical interpretations, while type IV$_0$ is known to arise semi-classically (in renormalized expectation values of standard stress-energy tensors).  In contrast type III$_0$ stands out in that it  
has neither a simple classical interpretation, nor even a simple semi-classical interpretation. 
We will also consider the robustness of this classification considering the stability of the different Hawking--Ellis types under perturbations. We argue that types II and III are definitively unstable, whereas types I and IV are stable.

\vspace{1pc}
\end{abstract}


\maketitle

\def\tr{{\mathrm{tr}}}
\def\diag{{\mathrm{diag}}}
\def\cof{{\mathrm{cof}}}
\def\pdet{{\mathrm{pdet}}}

\section{Introduction}

The classification of stress-energy tensors popularized by Hawking and Ellis~\cite{Hell} (which was in turn based on earlier work by Segre~\cite{Segre} and Pleba\'nski~\cite{Plebanski:1964}) has become an essential standard tool in developing more-or-less model-independent analyses of the implications of the Einstein field equations in astrophysical and cosmological contexts. (See for example references~\cite{wormholes, LNP, Martin-Moruno:2017}.) The Hawking--Ellis classification in turn feeds into (and to a large extent underlies) the formulation of the classical energy conditions~\cite{Hell}, and their non-linear and semi-classical quantum generalizations. (See for example references~\cite{wormholes, LNP, Martin-Moruno:2017, Martin-Moruno:2013a,Martin-Moruno:2013b,Martin-Moruno:2015,Visser:1994,Visser:1996a,Visser:1996b,Visser:1997}, and~\cite{Balakrishnan:2017,Akers:2017,Fu:2017,Fu:2016,Koeller:2015,Bousso:2015}, and related work on the Rainich conditions~\cite{Martin-Moruno:2017,1925,Misner:1957,Witten:1959,Senovilla:2000,Bergqvist:2001,Senovilla:2002,Bergqvist:2004,Plebanski:1994,Torre:2013,Krongos:2015,Santos:2016,Balfagon:2007}.) 

Note that in setting up the Hawking--Ellis classification,  we are working in a local orthonormal frame with $\eta^{ab} = \diag(-1,1,1,1)$ and looking for Lorentz-invariant eigenvalues and Lorentz-covariant eigenvectors:
\begin{equation}
(T^{ab}-\lambda \,\eta^{ab}) V_b = 0.
\end{equation}
(This is what mathematicians would call a ``generalized eigenvector problem''~\cite{HJ1,HJ2}.)

Equivalently one could raise and lower indices
\begin{equation}
(T^{a}{}_{b}-\lambda \,\delta^{a}{}_{b}) V^b = 0.
\end{equation}
(This is what mathematicians would call an ``ordinary eigenvector problem'', but it is the fact that $T^{a}{}_{b}$ is now generically \emph{not symmetric} that renders the classification programme mathematically non-trivial~\cite{HJ1,HJ2}.)
We will use $\sim_{\hbox{\tiny L}}$ to denote similarity under Lorentz transformations; and $\sim$ to denote similarity under generic non-singular transformations. 
Note that it is the Lorentz transformations that are used in the Hawking--Ellis classification, which is based on diagonalizing the stress-energy tensor (as much as possible) in a physical orthonormal basis, with one timelike and three spacelike basis vectors. In contrast, in order to get the Jordan normal form, one instead considers generic non-singular (possibly complex) transformations. The Jordan form is particularly useful in the classification based on the minimal polynomials~\cite{Martin-Moruno:2017}. However, whenever the basis in which the stress-energy tensor is expressed in its Jordan normal form does not contain a timelike eigenvector, (that is, it cannot be co-moving with any physical observer), then the Jordan form does not provide us with anywhere near so clear a physical interpretation. We will (mostly) work in (3+1) signature.

The basic idea we describe below is to ``simplify'' the Hawking--Ellis type I, II, III, IV classification as much as possible by isolating what we shall call the ``essential core'' of the type II, type III, and type IV stress-energy tensors; this being done by subtracting (special cases of) type I to simplify the (Lorentz invariant) eigenvalue structure as much as possible \emph{without disturbing the eigenvector structure}.
We shall see that these ``simplified cores'' have very nice and simple algebraic properties; and very straightforward mathematical characterizations.
Furthermore, physically types I and II$_0$ have very simple classical interpretations, while type IV$_0$ is known to arise semi-classically (in renormalized expectation values of standard stress-energy tensors).  In contrast type III$_0$ stands out in that physically it  
has neither a simple classical interpretation, nor even a simple semi-classical interpretation. 
Because of this, we shall spend some extra effort analyzing type III$_0$. 

\section{Essential core of the Hawking--Ellis classification}

Let us consider the Hawking--Ellis types I, II, III, and IV~\cite{Hell}.
(This classification is discussed in many places, including~\cite{wormholes, LNP, Martin-Moruno:2017, Martin-Moruno:2013a,Martin-Moruno:2013b,Martin-Moruno:2015,Visser:1994,Visser:1996a,Visser:1996b,Visser:1997}.) 
\begin{description}
\item[type I: ] Under a Lorentz transformation we can set
\begin{equation}
T^{\mu\nu} \sim_{\hbox{\tiny L}} \left[\begin{array}{c|ccc} 
\rho&0 &0 &0\\ \hline 0 &p_1& 0 &0\\  0&0& p_2 & 0\\0&0&0&p_3\\\end{array}\right];
\qquad
T^\mu{}_\nu \sim_{\hbox{\tiny L}}\left[\begin{array}{c|ccc} 
-\rho&0 &0 &0\\ \hline 0 &p_1& 0 &0\\  0&0& p_2 & 0\\0&0&0&p_3\\\end{array}\right].
\end{equation}
As in this case the stress-energy tensor is fully diagonalizable, one obtains the same matrix by expressing this tensor in the orthonormal basis formed by its eigenvectors
\begin{equation}
T^\mu{}_\nu \sim\left[\begin{array}{c|ccc} 
-\rho&0 &0 &0\\ \hline 0 &p_1& 0 &0\\  0&0& p_2 & 0\\0&0&0&p_3\\\end{array}\right].
\end{equation}
The eigenvalues are $\{-\rho,p_1,p_2,p_3\}$. This is as simple as type I gets. In (3+1) dimensions type I is invariantly characterized by the existence of a unique timelike eigenvector, implying the existence of three spacelike eigenvectors. (In Euclidean signature, since $\eta^{ab} \to \delta^{ab}$, all stress-energy tensors are type I.)

\item[type II:] Under a Lorentz transformation we can set
\begin{eqnarray}
\hspace{-2cm}
T^{\mu\nu} \sim_{\hbox{\tiny L}} \left[\begin{array}{cc|cc} 
\mu+f&f &0 &0\\ f &-\mu+f& 0 &0\\ \hline 0&0& p_2 & 0\\0&0&0&p_3\\\end{array}\right];
\,
T^\mu{}_\nu \sim_{\hbox{\tiny L}} \left[\begin{array}{cc|cc} 
-\mu-f&f &0 &0\\ -f &-\mu+f& 0 &0\\ \hline 0&0& p_2 & 0\\0&0&0&p_3\\\end{array}\right];
\end{eqnarray}
while under a generic non-singular similarity transformation we can set
\begin{equation}
T^\mu{}_\nu \sim\left[\begin{array}{cc|cc} 
-\mu&1 &0 &0\\ 0 &-\mu& 0 &0\\ \hline 0&0& p_2 & 0\\0&0&0&p_3\\\end{array}\right].
\end{equation}
The eigenvalues are $\{-\mu,-\mu,p_2,p_3\}$.
The non-trivial structure of the Jordan form is due to the null eigenvector associated to the double eigenvalue. That is, it is not expressed in a basis with a timelike eigenvector, though there are two spacelike eigenvectors associated to the $p_i$ eigenvalues.  As a matrix, type II is \emph{defective}, there are only three eigenvectors, not four.
Now from type II subtract out as much of type I as possible; and call what is left type II$_0$. 
(Make the eigenvalues as simple as possible; but do not disturb the eigenvectors.) 

Consider this:
\begin{equation}
(T^{\mu\nu})_{II_0} \sim_{\hbox{\tiny L}} \left[\begin{array}{cc|cc} 
f&f &0 &0\\ f &f& 0 &0\\ \hline 0&0&0 & 0\\0&0&0&0\\\end{array}\right];
\quad
(T^\mu{}_\nu)_{II_0}\sim_{\hbox{\tiny L}} \left[\begin{array}{cc|cc} 
-f&f &0 &0\\ -f &f& 0 &0\\ \hline 0&0&0 & 0\\0&0&0&0\\\end{array}\right];
\end{equation}
and
\begin{equation}
(T^\mu{}_\nu)_{II_0} \sim\left[\begin{array}{cc|cc} 
0&1 &0 &0\\ 0 &0& 0 &0\\ \hline 0&0& 0& 0\\0&0&0&0\\\end{array}\right].
\end{equation}
The eigenvalues are now as simple as possible, $\{0,0,0,0\}$, and there is still one null  and two spacelike eigenvectors.
Observe that type II$_0$ can be invariantly characterized by  the equation $[(T^a{}_b)_{II_0} ]^2 = 0$. (So it is nilpotent of order 2.)
We can also write $(T^{ab})_{II_0}  = f \ell^a \ell^b$ where $\ell$ is a null vector with unit time component. 
On the other hand, type II$_0$ tensors have a generalized timelike eigenvector, such that $[(T^a{}_b)_{II_0} ]t^b=f\ell^a$, implying $ [(T^a{}_b)_{II_0}]^2 t^b = 0$.
The minimum dimensionality for a type II$_0$ stress-energy tensor is (1+1).

\item[type III:] Under a Lorentz transformation we can set
\begin{equation}
T^{\mu\nu} \sim_{\hbox{\tiny L}} \left[\begin{array}{ccc|c} 
\rho&f &0&0\\ f &-\rho& f &0\\  0&f& -\rho & 0\\   \hline 0&0&0&p_3\\\end{array}\right];
\quad
T^\mu{}_\nu \sim_{\hbox{\tiny L}} \left[\begin{array}{ccc|c} 
-\rho&f &0&0\\ -f &-\rho& f &0\\  0&f& -\rho & 0\\   \hline 0&0&0&p_3\\\end{array}\right];
\end{equation}
where the parameter $f$ is unnecessarily set to 1 in reference~\cite{Hell}. The Jordan normal form of this tensor is
\begin{equation}
T^\mu{}_\nu \sim\left[\begin{array}{ccc|c} 
-\rho&1 &0 &0\\ 0 &-\rho& 1 &0\\  0&0& -\rho & 0\\ \hline 0&0&0&p_3\\\end{array}\right].
\end{equation}
The eigenvalues are $\{-\rho,-\rho,-\rho,p_3\}$, and there is a single null eigenvector associated to the triple eigenvalue, plus a single spacelike eigenvector associated to the eigenvalue $p_3$. As a matrix type III is \emph{defective}, there are only two eigenvectors, not four.

Now from type III subtract out as much of type I as possible; call what is left type III$_0$.
(Make the eigenvalues as simple as possible; but do not disturb the eigenvectors.) 
Consider this:
\begin{equation}
(T^{\mu\nu})_{III_0} \sim_{\hbox{\tiny L}} \left[\begin{array}{ccc|c} 
0&f &0&0\\ f &0& f &0\\  0&f&0& 0\\   \hline 0&0&0&0\\\end{array}\right];
\quad
(T^\mu{}_\nu)_{III_0} \sim_{\hbox{\tiny L}} \left[\begin{array}{ccc|c} 
0&f &0&0\\ -f &0& f &0\\  0&f&0& 0\\   \hline 0&0&0&0\\\end{array}\right];
\end{equation}
with Jordan form
\begin{equation}
(T^\mu{}_\nu)_{III_0} \sim\left[\begin{array}{ccc|c} 
0&1 &0 &0\\ 0 &0& 1 &0\\  0&0& 0& 0\\ \hline 0&0&0&0\\\end{array}\right].
\end{equation}
The eigenvalues are now as simple as possible, $\{0,0,0,0\}$, and we still have one null and one spacelike eigenvector.
Observe that type III$_0$ can be invariantly characterized by the equation $[(T^a{}_b)_{III_0} ]^3 = 0$, so that type III$_0$ tensors are nilpotent of order 3.

We can also write $(T^{ab})_{III_0}  = f (\ell^a s^b+s^a\ell^b)$; 
where $\ell$ is a null vector, and $s$ is spacelike and orthogonal to $\ell$.
For example, take $\ell^a \sim_{\hbox{\tiny L}} (1,0,1,0)$, and $s^a \sim_{\hbox{\tiny L}}(0,1,0,0)$.
To verify the invariant characterization $[(T^a{}_b)_{III_0} ]^3 = 0$, one can first check that $(\ell^a s^b+s^a\ell^b) \eta_{bc} (\ell^c s^d+s^c\ell^d) = \ell^a \ell^d$.
This implies that $[(T^a{}_b)_{III_0} ]^2 = f^2\; \ell^a \ell_b \in (T^a{}_b)_{II_0}$ and, therefore,  the square of a type III$_0$ tensor is a type II$_0$ tensor. 
Note that as $\ell$ is a null vector and $s$ is orthogonal to $\ell$, one obtains $[(T^a{}_b)_{III_0} ]^3 =  0.$

From the above, the minimum dimensionality for type III is (2+1).
In a certain technical sense type III$_0$ is a ``square root'' of type II$_0$,
but to take the ``square root'' one (at a minimum) has to go to one higher 
dimension than was needed for type II.
(This is vaguely similar to what happens for real $\to$ complex.)
Observe that the vector $\ell$ is the (unique) null eigenvector, since $[(T^a{}_b)_{III_0} ] \ell^b = 0$. 
In contrast the vector $s$ is not an eigenvector, since $[(T^a{}_b)_{III_0} ] s^b = f \ell^a$.
However $s$ is a generalized (spacelike) eigenvector of order 2, that is $[(T^a{}_b)_{III_0} ][(T^b{}_c)_{III_0} ] s^c = 0$. 
Moreover, for the generalized (timelike) eigenvector of order 3, one has $[(T^a{}_b)_{III_0} ] t^b = f s^a$, $[(T^a{}_b)_{III_0} ]^2 t^b = f^2\ell^a$, and $[(T^a{}_b)_{III_0} ]^3 t^b = 0$.

\item[type IV:] Under a Lorentz transformation we can set
\begin{equation}
T^{\mu\nu} \sim_{\hbox{\tiny L}} \left[\begin{array}{cc|cc} 
\rho&f &0 &0\\ f &-\rho& 0 &0\\ \hline 0&0& p_2 & 0\\0&0&0&p_3\\\end{array}\right];
\quad
T^\mu{}_\nu \sim_{\hbox{\tiny L}} \left[\begin{array}{cc|cc} 
-\rho&f &0 &0\\ -f &-\rho& 0 &0\\ \hline 0&0& p_2 & 0\\0&0&0&p_3\\\end{array}\right];
\end{equation}
while under a generic non-singular similarity transformation one can set
\begin{equation}
T^\mu{}_\nu \sim\left[\begin{array}{cc|cc} 
-\rho+if&0 &0 &0\\ 0 &-\rho-if& 0 &0\\ \hline 0&0& p_2 & 0\\0&0&0&p_3\\\end{array}\right].
\end{equation}
The eigenvalues are $\{-\rho+if ,-\rho-if,p_2,p_3\}$, and there are no causal eigenvectors; there are only two spacelike eigenvectors associated with the $p_i$ eigenvalues. (There are also two complex eigenvectors, that do not fit into the timelike/null/spacelike classification. The matrix is diagonalizable, but there are only two real eigenvectors, not four.)

Now from type IV subtract out as much of type I as possible; call what is left type IV$_0$.
(Make the eigenvalues as simple as possible; but do not disturb the eigenvectors.)
Since the eigenvalues of type I are all real, we will not be able to disturb the imaginary parts of the type IV eigenvalues. 
Consider this:
\begin{equation}
(T^{\mu\nu})_{IV_0} \sim_{\hbox{\tiny L}} \left[\begin{array}{cc|cc} 
0&f &0 &0\\ f &0& 0 &0\\ \hline 0&0& 0 & 0\\0&0&0&0\\\end{array}\right];
\quad
(T^\mu{}_\nu)_{IV_0} \sim_{\hbox{\tiny L}} \left[\begin{array}{cc|cc} 
0&f &0 &0\\ -f &0& 0 &0\\ \hline 0&0& 0 & 0\\0&0&0&0\\\end{array}\right];
\end{equation}
with Jordan form
\begin{equation}
(T^\mu{}_\nu)_{IV_0} \sim\left[\begin{array}{cc|cc} 
+if&0 &0 &0\\ 0 &-if& 0 &0\\ \hline 0&0& 0& 0\\0&0&0&0\\\end{array}\right].
\end{equation}
The eigenvalues are now $\{+if ,-if,0,0\}$.
Observe that a type IV$_0$ stress-energy tensor can be invariantly characterized either as
$([(T^\bullet{}_\bullet)_{IV_0} ]^2)^a{}_b = -f^2 [I_2 \oplus 0_2]^a{}_b$, or, if one prefers, as
$[T_{IV_0}] \eta^{-1} [T_{IV_0}] = -f^2 [\eta_2 \oplus 0_2]$.

We can also write $(T^{ab})_{IV_0}  = f (t^a s^b+s^at^b)$; 
were $t$ is a timelike vector, and the vector $s$ is spacelike and orthogonal to $t$.
For example, take $t^a \sim_{\hbox{\tiny L}} (1,0,0,0)$, and $s^a \sim_{\hbox{\tiny L}}(0,1,0,0)$. Then, it is easy to see that $([(T^\bullet{}_\bullet)_{IV_0} ]^2)^a{}_b=f^2(t^at_b-s^as_b)$ and, therefore,  $[(T^a{}_b)_{IV_0}]^3=-f^2(T^a{}_b)_{IV_0}$, as happens with pure imaginary numbers. So, in a vague sense type ${IV_0}$ is the imaginary version of type I.
The minimum dimensionality for type IV$_0$ is (1+1).
\end{description}

\section{General lessons}

What general lessons can we draw?
\begin{itemize}
\item 
Type I is the most generic case, corresponding to perfect fluids and anisotropic perfect fluids --- and even anisotropic solids. Any stress-energy tensor that has a ``natural rest frame'' (meaning a timelike eigenvector) is of type I.

\item  
Type II$_0$ corresponds to pure (coherent) radiation travelling at the speed of light.

\item  
Type III$_0$ corresponds no known source of stress-energy, neither classical nor quantum.

\item
Type IV$_0$ corresponds no known source of classical stress-energy, though in the semiclassical quantum realm renormalized expectation values of the stress-energy tensor are often of type IV$_0$. See for instance references~\cite{LNP, Martin-Moruno:2017, Martin-Moruno:2013a, Martin-Moruno:2013b, Martin-Moruno:2015}.

\item 
Note that all the essential core types, II$_0$, III$_0$, and IV$_0$, are traceless, $T^a{}_a = 0$.
\item
Types II$_0$ and  III$_0$ have $\det(T^a{}_b)=0$, whereas in contrast for type IV$_0$ we see that $\det(T^a{}_b)=f^2 >0$. 
\item 
Types II$_0$ and  III$_0$ have $\tr(T^2) = (T^a{}_b T^b{}_a)=0$, whereas for type IV$_0$ one has $\tr(T^2) = (T^a{}_b T^b{}_a)=-2f^2 <0$. 
\item
The absolutely simplest form of type III requires at least (2+1) dimensions and corresponds to
\begin{equation}
\hspace{-3cm}
(T^{\mu\nu})_{III_0} \sim_{\hbox{\tiny L}} \left[\begin{array}{ccc} 
0&f &0\\ f &0& f \\  0&f&0\\
\end{array}\right];
\,
(T^\mu{}_\nu)_{III_0} \sim_{\hbox{\tiny L}} \left[\begin{array}{ccc} 
0&f &0\\ -f &0& f \\  0&f&0\\
\end{array}\right];
\,
(T^\mu{}_\nu)_{III_0} \sim\left[\begin{array}{ccc} 
0&1 &0 \\ 0 &0& 1 \\  0&0& 0\\ 
\end{array}\right],
\end{equation}
with eigenvalues: $\{0,0,0\}$.
Since the simplest forms of types I, II, and IV can be defined in (1+1) dimensions, we see that type III stands out in this regard. (This implies, for instance that type III is incompatible with either planar or spherical symmetry. See for instance references~\cite{LNP,Martin-Moruno:2017, Martin-Moruno:2013a, Martin-Moruno:2013b, Martin-Moruno:2015}. 

\item According to the classification based on considering the minimal polynomial \cite{Martin-Moruno:2017}, type II$_0$ is degree 2.I (meaning that the minimal polynomial has degree 2 and the matrix has only one distinct eigenvalue). Similarly  type III$_0$ is degree 3.I, and type IV$_0$ is degree 3.III.

 \item When converting type II, III, IV stess-energy tensors to their essential cores,
    it is perhaps useful to note that one does not have to subtract the most general form of type I.
    The type I piece being subtracted is degenerate, in that it has either 2 or 3 of its Lorentz-invariant
    eigenvalues equal to the eigenvalue of its timelike eigenvector.
    
\end{itemize}

\section{Energy conditions}

\subsection{Type I}

The energy conditions for type I stress-energy are discussed in very many places, 
from the well-known Hawking--Ellis text~\cite{Hell}, to many secondary sources. (See for instance references~\cite{wormholes, LNP, Martin-Moruno:2017, Martin-Moruno:2013a,Martin-Moruno:2013b,Martin-Moruno:2015,Visser:1994,Visser:1996a,Visser:1996b,Visser:1997}). 
With regards to type I, we have nothing substantial to add. 

\subsection{Type II$_{0}$}

For type II$_{0}$ stress-energy we have $(T^{ab})_{II_0}=f \ell^a \ell^b$. For the standard energy conditions (NEC, WEC, SEC, DEC) this implies the straightforward results:
\begin{itemize}
\item The Null Energy Condition (NEC) is satisfied iff $f\geq0$.
\item The Weak Energy Condition (WEC) is satisfied iff $f\geq0$.
\item Since $T=0$, the Strong Energy Condition (SEC) is satisfied iff $f\geq0$.
\item The Dominant Energy Condition (DEC) is satisfied iff $f\geq0$.\\ (Note that DEC=WEC+FEC, and see comments on the Flux Energy Condition (FEC) below.)
\end{itemize}
For the less standard energy conditions (FEC, TEC, TOSEC, DETEC  \cite{Martin-Moruno:2013b}) we have:
\begin{itemize}
\item FEC (Flux Energy Condition) is marginally satisfied. \\
For any timelike observer $V$, the flux $F^a = T^{ab} V_b = (f \ell^b V_b) \ell^a$ is always null.
\item
TEC (Trace Energy Condition) is marginally satisfied ($T=0$).
\item 
TOSEC (Trace-of-square energy condition) is marginally satisfied ($\tr[T^2]=0$).
\item 
DETEC (Determinant energy condition) is marginally satisfied  ($\det[T]=0$).
\end{itemize}

\subsection{Type III$_{0}$}

For type III$_0$, that is $(T^{ab})_{III_0}  = f (\ell^a s^b+s^a\ell^b)$, the situation becomes more subtle. For the standard energy conditions:
\begin{itemize}
\item
NEC is violated.\\
For any null vector $k$ we have $T_{ab} k^a k^b = 2 f (k\cdot\ell)(k\cdot s)$.
We can always choose signs to enforce $k\cdot\ell\geq 0$, but $k\cdot s$ can easily flip sign.
Thus type III$_0$ cannot satisfy the inequality on which the NEC is based for all null vectors.

\item
WEC is violated.\\
For any timelike vector $V$ we have $T_{ab} V^a V^b = 2 f (V\cdot\ell)(V\cdot s)$.
We can always choose signs to enforce $V\cdot\ell> 0$, but $V\cdot s$ can easily flip sign.
Thus type III$_0$ cannot satisfy the the inequality on which WEC is based for all timelike vectors.

\item
SEC is violated.\\
SEC = WEC because $T=0$.

\item
DEC is violated. 
\\
(DEC=WEC+FEC, and FEC is violated; see below.)

\end{itemize}
For the less-standard energy conditions:
\begin{itemize}
\item FEC is violated. \\
For timelike 4-velocities $V$, the observed flux is 
\begin{equation}
F_a = T_{ab} V^b = f [s_a (\ell_b V^b) +\ell_a (s_b V^b)].
\end{equation}
But then $F_aF^a  = f^2(\ell_b V^b)^2 > 0$, so the flux vector is always spacelike,
indicating that type III$_0$ corresponds to \emph{tachyonic matter}.
\item
TEC is marginally satisfied ($T=0$).
\item 
TOSEC is marginally satisfied ($\tr[T^2]=0$).
\item 
DETEC is marginally satisfied ($\det[T]=0$).
\end{itemize}

\subsection{Type IV$_{0}$}

For type IIV$_0$, $(T^{ab})_{IV_0}  = f (t^a s^b+s^at^b)$. So, for the standard energy conditions:
\begin{itemize}
\item 
NEC is violated.\\
For any null vector $k^a$ we have $T_{ab} k^a k^b=2f(t\cdot k)(s\cdot k)$.
We can always choose signs so that $(t\cdot k)>0$, but $(s\cdot k)$ can easily flip sign.
Thus type IV$_0$ cannot satisfy the the inequality on which NEC is based for all null vectors.

\item
WEC is violated.\\
For any timelike vector $V^a$ we have $T_{ab} V^a V^b=2f(t\cdot V)(s\cdot V)$.
We can always choose signs so that $(t\cdot V)>0$, but $(s\cdot V)$ can easily flip sign.
Thus type IV$_0$ cannot satisfy the inequality on which WEC is based for all timelike vectors.

\item 
SEC is violated.\\
SEC = WEC because $T=0$.

\item
DEC is violated.
\\
(DEC=WEC+FEC, and FEC is violated; see below.)

\end{itemize}
For the less-standard energy conditions:
\begin{itemize}
\item FEC is violated. \\
For any timelike 4-velocity $V$ we have the flux vector
\begin{equation}
F_a = T_{ab} V^b = f [t_a (s \cdot V) +s_a (t \cdot V)].
\end{equation}
But then $F_aF^a  = f^2[(t \cdot V)^2 -(s \cdot V)^2]> 0$ where the last step takes into account that the projection of a timelike vector along a timelike direction is larger than its projection along a spacelike direction. So the flux vector is always spacelike,
indicating that type IV$_0$ corresponds to \emph{tachyonic matter}.

\item
TEC is marginally satisfied ($T=0$).
\item 
TOSEC is violated ($\tr[T^2]=-2f^2<0$).
\item 
DETEC is satisfied ($\det[T]=f^2>0$).
\end{itemize}

\section{Stability of the Hawking--Ellis types}

Now let us consider how stable the Hawking--Ellis types are under infinitesimal perturbations. 
One reason for being particularly interested in this is to understand the potential pitfalls of relying on numerical estimates and calculations of semi-classical renormalized stress-energy tensors; there will always be numerical approximation and round-off issues, and we would like to understand how these issues affect the Hawking--Ellis classification. 

We find it useful to first step back to consider standard purely mathematical matrix results regarding diagonalizable versus defective matrices~\cite{HJ1,HJ2}:
\vspace{-7mm}
\begin{quote}
\item Generic perturbations of any matrix, (regardless of whether the original matrix is diagonalizable under similarity transformations or not), will lead to a diagonalizable matrix. 
\end{quote}
The point is that generic perturbations will lift any eigenvalue degeneracy which might be present, while preserving or inducing an eigenvalue degeneracy would require a very non-generic perturbation~\cite{HJ1,HJ2}. Since distinct eigenvalues are a sufficient condition for diagonalizability, generic perturbations of any matrix will lead to a diagonalizable (under similarity transformations) matrix --- any matrix is infinitesimally close to a diagonalizable matrix~\cite{HJ1,HJ2}.

Within the context of the Hawking--Ellis classification, the diagonalizable stress-energy tensors correspond to types I and IV, whereas types II and III are non-diagonalizable.  
So we have:
\begin{itemize}
\item 
Perturbing generic type I, (with all eigenvalues unequal), generically leads to type~I.
\item
Perturbing degenerate type I, (with some eigenvalues equal), generically leads to either type~I or  type~IV. 
\item
Perturbing type II  generically leads to either type~I or  type~IV.
\item
Perturbing type III  generically leads to either type~I or  type~IV.
\item
Perturbing type IV generically leads to type IV.
\end{itemize}
To clarify this point further, note that for all of the Hawking--Ellis types there is always an orthonormal basis where any stress-energy tensor can be written as
\begin{equation}
T^\mu{}_\nu\sim \left[\begin{array}{ccc|c} \label{Tp}
-\rho&f_1 &f_2 &0\\ -f_1 &p_1& s &0\\  -f_2&s& p_2 & 0\\ \hline 0&0&0&p_3\\\end{array}\right].
\end{equation}
The characteristic polynomial of this tensor is of the form
\begin{equation}
c(\lambda)=(p_3-\lambda)[\lambda^3+b\lambda^2+c\lambda+d\,].
\end{equation}
Apart from the eigenvalue $\lambda=p_3$, the multiplicity of the other eigenvalues (which is closely related to the Hawking--Ellis type of the stress-energy tensor) will depend on the roots of the cubic equation resulting from equating to zero the square bracket of the characteristic polynomial.
Defining the characteristic $\Delta = 18 bcd-4b^3d+b^2c^2-4c^3-27d^2$ of the cubic polynomial,
those roots can be classified as follows:
\begin{itemize}
\item 
$\Delta>0$: There are 3 distinct real roots; thus, we have type I stress-energy.
\item 
$\Delta=0$: There is at least one multiple eigenvalue; we have either degenerate type I stress energy, or alternatively type II or type III stress-energy.
\item 
$\Delta<0$: There is 1 real root and 2 complex roots; we have  type IV stress-energy.
\end{itemize}
A small perturbation of the stress-energy tensor implies a small perturbation of the value of $\Delta$, disturbing its value from zero if it is initially zero. Thus, degenerate type I, and types II and III, are indeed unstable.

Numerical calculations of renormalized stress-energy tensors in spherical symmetry often lead to thick spherical regions of type I and (in the Unruh quantum vacuum state) thick spherical regions of type IV, see references~\cite{LNP, Martin-Moruno:2017,Martin-Moruno:2013a, Martin-Moruno:2013b,Martin-Moruno:2015,Visser:1996a, Visser:1996b, Visser:1997}. From the above so far very general discussion, such regions must be separated by a region of either degenerate type I, or type II, but since that region is numerically unstable, it can at best be a thin zero-thickness shell whose location cannot be estimated to better than  numerical precision. 
But we can actually do better that this --- since type IV in spherical symmetry requires $f=T_{\hat t \hat r} \neq 0$,  (so that we must be in the Unruh quantum vacuum state, and conservation of stress energy then implies $f\neq0$ everywhere throughout the spacetime),  we can write the interesting part of the stress-energy tensor for any spherically symmetric scenario as 
\begin{equation}
(T^{\mu\nu}) = 
\left[\begin{array}{cc} \rho &f \\ f & p \end{array}\right] 
= 
\left[\begin{array}{cc} f+\mu+\delta &f \\ f & f-\mu+\delta \end{array}\right].
\end{equation}
The Lorentz invariant eigenvalues are easily determined to be $-\mu\pm\sqrt{\delta(\delta+2f)}$. Then the region where $\delta(\delta+2f)>0$ is type I, while the region where $\delta(\delta+2f)<0$ is type~IV. At the transition layer, we have either $\delta=0$ or $\delta=-2f$.  This corresponds to
\begin{equation}
(T^{\mu\nu}) = 
\left[\begin{array}{cc} \mu\pm f &f \\ f & -\mu \pm f \end{array}\right],
\end{equation}
both of which are type II. So the transition layer is guaranteed to be type II.
(This is indeed what seems to happen in practice.)

Particular examples of perturbations of degenerate type I, type II, and type III are most effectively dealt with by directly applying perturbation arguments within the context of the type II$_0$ and type III$_0$ classifications.
\begin{itemize}
\item 
Consider this perturbation $\epsilon$ of a degenerate type I stress-energy tensor
\begin{equation}
\hspace{-2.4cm}
(T^{\mu\nu}) \sim_{\hbox{\tiny L}} \left[\begin{array}{cc|cc} 
\rho+\epsilon &0 &0 & 0\\ 0 &-\rho+\epsilon & 0 &0\\ \hline 0&0&p_2 & 0\\0&0&0&p_3\\\end{array}\right];
\,
(T^\mu{}_\nu)\sim_{\hbox{\tiny L}} \left[\begin{array}{cc|cc} 
-\rho-\epsilon&0&0 &0\\ 0&-\rho+\epsilon &0& 0 \\ \hline 0&0&p_2&0\\0&0&0&p_3\\\end{array}\right].
\end{equation}
For $\epsilon=0$ this is degenerate type I. For $\epsilon\neq 0$ this is type I. The eigenvalues are $\{\rho\pm\epsilon,0,0\}$. 
Thus perturbations of degenerate type I can easily lead to  type I. 

\item 
Consider this  (different) perturbation $\epsilon$ of a degenerate type I stress-energy tensor
\begin{equation}
\hspace{-1cm}
(T^{\mu\nu}) \sim_{\hbox{\tiny L}} \left[\begin{array}{cc|cc} 
\rho &\epsilon &0 & 0\\ \epsilon &-\rho & 0 &0\\ \hline 0&0&p_2 & 0\\0&0&0&p_3\\\end{array}\right];
\quad
(T^\mu{}_\nu)\sim_{\hbox{\tiny L}} \left[\begin{array}{cc|cc} 
-\rho&\epsilon&0 &0\\ -\epsilon&-\rho &0& 0 \\ \hline 0&0&p_2&0\\0&0&0&p_3\\\end{array}\right].
\end{equation}
For $\epsilon=0$ this is degenerate type I. For $\epsilon\neq 0$ this is type IV. The eigenvalues are $\{\rho\pm i\epsilon,0,0\}$. 
Thus perturbations of degenerate type I can easily lead to  type IV.

\item 
Consider this perturbation $\epsilon$ of a type II$_0$ stress-energy tensor
\begin{equation}
\hspace{-1.5cm}
(T^{\mu\nu}) \sim_{\hbox{\tiny L}} \left[\begin{array}{cc|cc} 
f&f+\epsilon &0 &0\\ f+\epsilon &f& 0 &0\\ \hline 0&0&0 & 0\\0&0&0&0\\\end{array}\right];
\,
(T^\mu{}_\nu)\sim_{\hbox{\tiny L}} \left[\begin{array}{cc|cc} 
-f&f +\epsilon&0 &0\\ -f-\epsilon &f& 0 &0\\ \hline 0&0&0 & 0\\0&0&0&0\\\end{array}\right].
\end{equation}
For $\epsilon=0$ this is type II$_0$. For $\epsilon\neq 0$ this is, (depending on the sign of $\epsilon f$),
either type I or type IV. The eigenvalues are $\{\pm\sqrt{-2\epsilon f} +O(\epsilon^{3/2}),0,0\}$. 
Thus perturbations of type II$_0$ (and so \emph{mutatis mutandi} type II) can lead to either type I or type IV.

\item
Consider finally this perturbation $\epsilon$ of a type III$_0$ stress-energy tensor
\begin{equation}
\hspace{-1.5cm}
(T^{\mu\nu})\sim_{\hbox{\tiny L}} \left[\begin{array}{ccc|c} 
0&f &0&0\\ f &0& f+\epsilon &0\\  0&f+\epsilon&0& 0\\   \hline 0&0&0&0\\\end{array}\right];
\,
(T^\mu{}_\nu)\sim_{\hbox{\tiny L}} \left[\begin{array}{ccc|c} 
0&f &0&0\\ -f &0& f+\epsilon &0\\  0&f+\epsilon&0& 0\\   \hline 0&0&0&0\\\end{array}\right].
\end{equation}
For $\epsilon=0$ this is type III$_0$. For $\epsilon\neq 0$ this is, (depending on the sign of $\epsilon f$),
either type I or type IV. The eigenvalues are $\{\pm\sqrt{+2\epsilon f} +O(\epsilon^{3/2}),0,0\}$. 
Thus perturbations of type III$_0$ (and so \emph{mutatis mutandi} type III) can lead to either type I or type IV. \end{itemize}
\enlargethispage{10pt}
Overall, we see that non-degenerate type I and type IV are stable under perturbations, while degenerate type I, and types II and III, are unstable.

\section{Discussion}

Note that all the essential core types, II$_0$, III$_0$, and IV$_0$ are significantly simpler to work with than the full type II, III, IV, stress-energy tensors, and have rather nice algebraic properties. Physically they correspond to subtracting as much of type I as possible, to always make the eigenvalues as simple as possible (while preserving eigenvector structure). 

We have also seen that the energy conditions are easier to deal with using  essential core types, II$_0$, III$_0$, and IV$_0$, and that it is easier to get a grasp of perturbative stability of the Hawking--Ellis classification using the essential core II$_0$ and III$_0$ types.

Furthermore, focussing on type III$_0$, (rather than the full type III), makes  it a little clearer just how physically odd type III really is. Type III stands out in that it does not seem to have any straightforward physical interpretation in either classical or quantum physics, a point we plan to address in future work~\cite{forthcoming}.

Finally, we have discussed the stability of the Hawking--Ellis classification under infinitesimal perturbations. We have shown that types II and III, which are those energy tensors that are non-diagonalizable, are unstable and generically decay into either type I or type IV. In contrast non-degenerate type I and generic type IV remain so under generic perturbations. Note that degenerate type I, which has a multiple eigenvalue but only one timelike eigenvector, is also unstable.

\section*{Acknowledgments}
PMM acknowledges financial support from the projects FIS2014-52837-P (Spanish MINECO) and FIS2016-78859-P (AEI/FEDER, UE).
MV acknowledges financial support via the Marsden Fund administered by the Royal Society of New Zealand.

\clearpage
\section*{References}

  
%
\end{document}